\documentclass[format=sigconf]{acmart}
\acmConference[ELS’21]{the 14th European Lisp Symposium}{May 03--04 2021}{Online, Everywhere}
\acmDOI{10.5281/zenodo.4713971}
\acmISBN{}
\setcopyright{rightsretained}
\copyrightyear{2021}

\usepackage{graphicx}
\usepackage{amsmath,amsthm}

\usepackage[finalizecache=false,frozencache=true]{minted}
\setminted{fontsize=\footnotesize}

\usepackage{enumitem}
\setlist[description]{leftmargin=\parindent,labelindent=0em}

\usepackage{multicol}
\usepackage{cleveref}

\usepackage{xspace}
\newcommand{\aether}{\texttt{aether}\xspace}
\newcommand{\register}{\texttt{register}\xspace}
\newcommand{\unregister}{\texttt{unregister}\xspace}
\newcommand{\send}{\texttt{send-message}\xspace}
\newcommand{\receive}{\texttt{receive-message}\xspace}

\newtheorem{theorem}{Theorem}
\theoremstyle{definition}
\theoremstyle{remark}
\newtheorem{remark}[theorem]{Remark}
\newtheorem{example}[theorem]{Example}

\begin{document}

\title{\aether: Distributed system emulation in Common Lisp}

\author{Eric C.\ Peterson}
\affiliation{
\institution{Eigenware}
\city{Berkeley}
\state{CA} 
\country{USA}}
\email{peterson.eric.c@gmail.com}

\author{Peter J.\ Karalekas}
\affiliation{
\institution{Eigenware}
\city{Berkeley}
\state{CA} 
\country{USA}}
\email{peter@karalekas.com}



\begin{abstract}
We describe a Common Lisp package suitable for the design, specification, simulation, and instrumentation of real-time distributed algorithms and hardware on which to run them.
We discuss various design decisions around the package structure, and we explore their consequences with small examples.
\end{abstract}

\maketitle

\section{Introduction}

In this paper, we consider the dual problems of modeling and designing a complex distributed system, at the levels of both hardware and software.
Those working in this problem space labor under an overwhelming number of variables: the structure and properties of the network, the structure and properties of the compute nodes on the network, the precise behavior of the software on the nodes, and so on.
These variables have complex interdependencies, and the relative sizes of their effects depend strongly on the ultimate application and its performance requirements.
Meanwhile, building a physical system to verify the performance of any collection of settings can be prohibitively expensive.
Thus, this problem is especially well-suited for simulation.

Common Lisp is exceptional at incrementally modeling complex systems whose full set of constraints is not known at start.
Its features allow one to nonuniformly increase the fidelity of the model in whatever areas make the most impact.
While Common Lisp is not initially well-suited specifically to distributed programming, it is both generally adaptable and organizable, so that one can make a principled account of the augmentations needed to support this style of programming and its simulation.
We offer here a Common Lisp framework for the design, programming, and analysis of distributed systems which are, in a variety of senses, not fully specified at the inception of the project.

To set the stage more concretely, let us discuss the specific application which drove the initial development of this framework.
In quantum computing, our domain of interest, the classical electronics which drive a typical system are a fleet of processors and programmable radio-frequency waveform generators and receivers.
These are highly concurrent systems which operate on micro- or nanosecond timescales, and they must solve coordination problems to determine the next step in a program.
Basic forms of coordination (e.g., time synchronization) are required for any kind of operation, but they are especially critical for quantum error correction~\cite{KitaevQCAEC,KitaevQEC}.
Syndrome decoding and error correction require active communication and joint processing of feedback from different areas of the quantum device in hard real-time.
Meanwhile, because the practice of quantum computing is so nascent and ``artisinal'', the details of the device, its steering electronics, and even the decoding scheme itself are all in flux---indeed, their design has the potential to be \emph{driven} by the predicted performance characteristics of a specified decoding procedure.

To analyze the behavior of such systems, it is valuable to have a flexible emulation framework available which can accommodate myriad hardware components with varying levels of detail:
\begin{description}
\item[Network]
The structure, protocol, and properties should all be programmable.
\item[Compute nodes]
The architecture of a particular node may be specified in complete detail, e.g., its behavior can be modeled by an instruction-by-instruction emulator with fully specified timing.
Similarly, the amount of computational power available to a node should be configurable.
\item[System composition]
In all aspects, it should be possible to work with heterogeneous components, e.g., different processors with different microarchitectures and different clock rates.
\item[Macroscopic defaults]
In all aspects, it should be possible to ignore implementation details by falling back on default behaviors.
These defaults may be unrealistic in a limited-resource regime (e.g., infinite network transmission capability), but they must nonetheless be fully specified.
\end{description}
We designed our software package, \aether~\cite{aether}, with these kinds of flexibility in mind: \aether provides support for the time-domain emulation of components with user-defined behavior, communication among these components, the steady offloading of execution semantics from standard Common Lisp to an application-specific DSL, the debugging of such software, and the analysis of performance effects in the forms of time, compute, and network requirements.

In sequels to this paper, we describe an application to topological quantum error correction, provide an implementation using \aether, and apply \aether's performance analysis features~\cite{PetersonKaralekasAnatevka}.

\begin{remark}
\aether is \emph{not} suitable for performant concurrent programming in Common Lisp, but this is not its goal.
There are many such libraries available, including \texttt{cl-actors}~\cite{clactors}, \texttt{cl-muproc}~\cite{clmuproc}, \texttt{erlangen}~\cite{erlangen}, \texttt{memento-mori}~\cite{mementomori}, and \texttt{simple-actors}~\cite{simpleactors}, though this list is in no way exhaustive.
Ours differs in that it intends to capture the performance features of distributed software running atop emulated hardware, whereas these others are concerned with performing the distributed computation itself as quickly as possible.
\end{remark}

\section{Many-component simulation}

The bedrock of \aether is in its simulation loop.
A key quality of a digital system is that its components operate against various clocks, where on each tick they enact some operation before delaying until the next tick.
Additionally, information passed through one component (e.g., a network interface or a memory bus) may arrive at some other component for reading after a possible delay.

The two key data types facilitating this kind of simulation are:
\begin{description}
    \item[\texttt{event}]
    A callback and a time at which to trigger it.  In addition to any other side effects, the callback returns a list of future events to reinject into the simulation.
    \item[\texttt{simulation}]
    A collection of events yet to be run and a time value, called the \textit{horizon}, which tracks the time before which history has been fully fixed.  In particular, no events in the collection may have trigger times before the horizon value.
\end{description}
The utility function \texttt{simulation-add-event} can be used to set up the initial contents of a simulation, and \texttt{simulation-run} can be used to process events in time order%
\footnote{%
Events scheduled for the same time are run serially, but not according to any convention.
It could be useful for debugging race conditions to expose a programmable hook when breaking such ties.
}
either until exhaustion or until a user-specified condition has been met (e.g., a timeout).

To facilitate the authoring of event callbacks, \aether provides the context macro \texttt{with-scheduling}, which executes its body in sequence while offering the following convenience forms to collect events to be returned at the close of the macro:%
\footnote{Compare SBCL's \texttt{define-vop}'s local macro \texttt{inst}.}
\begin{description}
    \item[\texttt{schedule}]
    Stashes an event for return at the close of the macro.
    \item[\texttt{schedule*}]
    Consumes the results of another callback and adjoins its events to the current list of events to be returned.
    \item[\texttt{finish-with-scheduling}]
    Exits \texttt{with-scheduling}, either with a specified return value or with the events collected thus far.
\end{description}

\begin{example}
In \Cref{DelayCallbackExample}, we illustrate using these constructs to generate a few callback events.

\begin{figure}[t]
\begin{minted}{cl}
(defun g (now)
  (with-scheduling (format t "g: ~a~%" now)))

(defun f (now)
  (with-scheduling
    (format t "f: ~a~%" now)
    (schedule #'g (+ now (/ (random 4) (1+ (random 4)))))
    (schedule #'f (+ now 1))))

(let ((simulation (make-simulation)))
  (simulation-add-event simulation (make-event :callback #'f))
  (simulation-run simulation :canary (canary-until 2))
  (values))

;; f: 0
;; f: 1
;; g: 5/4
;; g: 3/2
;; f: 2
\end{minted}
\caption{Example event callback \texttt{f} which triggers itself after a fixed delay and another function \texttt{g} after a random delay.}\label{DelayCallbackExample}
\end{figure}
\end{example}

Often, the components of a simulation are properly thought of as objects, each of which has a default behavior associated with it.
For example, a particular simulation component may be responsible for emulating the behavior of a processor, which on each tick decodes and enacts the instruction at its program counter.
To facilitate this, the generic function \texttt{define-object-handler} captures the default callback behavior of an object class, and when a simulation processes an event whose ``callback'' is an object, its callback is instead supplied by \texttt{define-object-handler}.

\begin{example}
In \Cref{DOHExample}, we define a processor emulator's heartbeat loop by using default object callbacks.

\begin{figure}[t]
\begin{minted}{cl}
(defstruct processor
  "State of a simple CPU."
  (data-stack      nil :type list)
  (program-counter 0   :type (integer 0))
  (instructions    nil :type vector :read-only t))

(define-object-handler ((self processor) now)
  ;; NOTE: passing the bare object "self" to schedule
  (schedule self (+ now (/ *processor-rate*)))
  (destructuring-bind (operator &rest arguments)
      (aref (processor-instructions self)
            (1- (processor-program-counter self)))
    (incf (processor-program-counter self))
    (ecase operator
      ;; example definitions of instruction behavior
      (halt  ; halts processor execution
       (finish-with-scheduling))
      (push  ; place constant on stack
       (push (first arguments) (processor-data-stack self)))
      (muli  ; stack-constant multiplication
       (push (* (pop (processor-data-stack self))
                (first arguments))
             (processor-data-stack self)))
      ;; ...
      )))
\end{minted}
\caption{\texttt{define-object-handler} used to describe the transition rule for a toy microprocessor.}\label{DOHExample}
\end{figure}
\end{example}

\begin{remark}
A simulation as described above can be neatly implemented via a priority queue.
However, such an implementation turns out to be under-performant in cases of interest: for large collections of processors with identical start times and identical clock frequencies, there are very few distinct key values in the priority queue.
For this reason, \aether's default implementation of a \texttt{simulation} uses a data structure with the interface of a priority queue but whose internals take advantage of repeated keys.
\end{remark}

\section{Messages and Message Passing}

While we have discussed some aspects of time-domain simulation, we have not made any particular reference to the communicative aspect of distributed computation.
We make the following fundamental assumption: typical communication channels are physically synchronous (although perhaps lossy), but their \emph{access} is subject to contention and hence appears as asynchronous.
Accordingly, \aether's networking capabilities are split into two halves: the simulation of a ``physical'' layer, which processes serially-supplied data against a constant clock; and primitives which communicate over this layer, which may appear to be asynchronous as delay accumulates in the simulation.
We discuss the two layers in turn.

The physical network is made up of \texttt{courier} objects.
As an interface to the communication API, each courier maintains a set of inboxes, indexed by address, which collect packets to be retrieved.
Additionally, each courier maintains a transmission queue of packets to be sent: if the destination address resides on this courier, the message is stashed in the appropriate inbox; if not, it is routed to the next courier on the transmission path.
The default courier class implements a network with all-to-all connectivity: every courier on the network is one hop away from every other.
We also provide a more specialized example implementation where the couriers are arranged in a grid, direct communication is only allowed between nearest neighbors, and the couriers route packets accordingly.%
\footnote{%
More explicitly, each destination address consists of a courier ID and an inbox ID registered at that courier.
In the gridded example, a courier's ID is its coordinate position in the grid, and couriers route packets according to logic such as: if the requested address's $x$--coordinate is to the left of this one, forward this packet to the courier to the immediate left, so that the routing table is procedurally generated.}
One informs \aether which courier services a block of code by binding the dynamic variable \texttt{*local-courier*}, which allows the code to interact with the networking layer.%
\footnote{In practice, this binding is done automatically.}

The communication API then provides a point of contact between this physical networking layer and the distributed programmer.
It defines the following primitives:
\begin{description}
\item[\register]
Opens a new inbox at the local courier and returns the key to that inbox.
\item[\send]
Sends a payload to a specified address.
Any message may be sent to any inbox, provided the address of that inbox is known.
\item[\receive]
Checks the indicated inbox for messages of a specified type.
If there is such a message, it binds it and executes an associated block of code.
If no message is found, the receive terminates and execution resumes as before.%
\footnote{%
More exactly, the body of \receive can carry different branches which specialize on different message types.
The inbox is first scanned for instances of the first message type.
By default, the entire inbox is considered, and if one or more messages is found, the indicated place is bound to the first (i.e., oldest) matching message, the associated block of code is executed, and then control exits from \receive without trying its other branches.
By toggling the \texttt{peruse-inbox?} argument, the user can tell it to only consider the first message in the inbox, rather than the entire inbox.
If no message of the first type is found, then it proceeds to search for messages of the second type.
Message processing continues in this way until all the message types are exhausted.
In the case of exhaustion, \receive executes an \texttt{otherwise} block if one is available, and then cedes control to its caller either way.}
The key associated to an inbox is required to receive messages from it.
\item[\unregister]
Releases an inbox resource from the courier.
A released inbox address can never again be used to retrieve messages.
\end{description}

\begin{example}
These primitives are enough to build various convenience methods.
For example, \Cref{BusyWaitEx} illustrates an implementation of a ``synchronous receive'', which waits to resume execution until a receive has been successful.%
\footnote{In the next section, we discuss a standard implementation within \aether of waiting for \receive to complete.}
\aether also provides \texttt{send-message-batch}, a utility which sends messages to a sequence of target addresses.

\begin{figure}[t]
\begin{minted}{cl}
(defun busy-wait (address continue-computation now)
  (with-scheduling
    (receive-message address message-place
      ;; if there's a done-message available,
      (done-message
        ;; pass it to continue-computation and finish
        (schedule* (funcall continue-computation message-place)))
      ;; otherwise, delay and loop
      (otherwise
        (schedule (a:curry #'busy-wait address continue-computation)
                  (+ now *busy-wait-delay*))))))
\end{minted}
\caption{An example callback that performs a busy-wait.}\label{BusyWaitEx}
\end{figure}
\end{example}

\begin{remark}
Message passing in \aether comes with significant guarantees: messages are always delivered, and messages sent from the same source to the same target are received in order.
Neglecting the effects of transmission delay (e.g., when working with a single courier), \aether acquires the semantics of the \textsf{LOCAL} model of distributed computing~\cite{Linial}.
With transmission and processing delays incorporated, \aether acquires the semantics of the \textsf{DECOUPLED} model~\cite{CDGFRR,DGFFR}.

Additionally, each message is optionally tagged with a ``reply address''.
This is available for user-defined behavior, but it is also used automatically by the networking layer in the case that the message is sent to an address which does not exist (e.g., if \unregister has since been called to close its destination inbox).
In this event, the destination courier will automatically generate a ``return to sender'' message and send it to the reply address so as to signal the failure.
\end{remark}

\begin{remark}
Neither the simulation framework nor the messaging framework provide a built-in mechanism for communicating with the outside world.
Instead, we leave it up to the user to decide how information should be extracted from the simulation.
Typical options include direct inspection of simulated objects or via a user-supplied object that converts simulation-internal messages to simulation-external side effects.
\end{remark}

\section{Actors and Processes}

A robust message-passing system opens the door for actor-based distributed computing~\cite{HBS}.
Each actor, which might be thought of as a physical processor or as a green thread on some ambient physical processor, is responsible for some component of a computation, which it completes by communicating with other actors through the network.

The basic notion of an actor is captured in \aether as a \texttt{process}.
In order to render invisible the difference between processes residing on the same ambient processor and those residing on different nodes in the network, processes ought not to modify each others' state directly.
Instead, each process registers an inbox with the courier to serve as its ``public'' address, and processes communicate with one another via these inboxes.
The processes then proceed to act at regular intervals, with the pace of each set by its \texttt{process-clock-rate}.
On each such tick, a process services inbound requests at its public address through user-defined handlers and makes progress on its own computations through user-defined procedures.

Service handlers are specified using \texttt{define-message-handler}, which takes as arguments the process (specialized by type), the message (specialized by type), and the current clock value.
The body of a handler might access or modify the process's state, cause it to jump to a new code segment to perform a computation, generate messages, and more.
The main constraint is that handlers function as ``interrupts'', in the sense that they are given a fixed amount of time (viz., one tick) to perform computation before they are required to terminate.
Apart from the individual specification of these handlers, \texttt{define-message-dispatch} specifies the ordered list of service handlers that a process employs to act on messages received at its public address (by repeatedly calling out to \receive).
Each handler is optionally guarded by a predicate that is computed from the current state of the process, so that services can be selectively enabled.

The computational trajectory of a process is captured by its ``command stack'', which is a stack of commands yet to be performed, each specified as a cell of shape \texttt{(COMMAND-NAME . ARGS)}.
On each tick, a process automatically pops the next command off of its stack and performs the specified action, whose behavior is set out by \texttt{define-process-upkeep}.
Typical command behaviors include:
\begin{description}
    \item[Local state modification]
    The command might modify the local state of the process, e.g., by changing some of its slot values.
    \item[Message-passing]
    The command might send new messages or draw waiting messages out of inboxes.
    \item[Intermediate computation]
    The command might perform intermediate computation, e.g., collating values received from different network sources.
    \item[Procedure continuation]
    The command might push more items onto the command stack (viz., using \texttt{process-continuation}).
    This has the effect of delaying the command that was to be processed immediately next, in favor of performing some other command (or sequence of commands) first.%
    \footnote{%
    For instance, one may implement a \texttt{while} loop by  testing the loop predicate and, conditionally on the test coming back positive, pushing the loop body and the \texttt{while} loop back onto the command stack.
    }
\end{description}

\noindent
In this way, the command stack simultaneously serves the purpose of a traditional call stack, a program counter, and executable memory.%
\footnote{
While not set in stone, this unusual design choice is a direct consequence of the flexibility we have elected for in process semantics.
In order to make use of a program counter to track program state, one needs the ability to jump (so, in particular, to return) to an arbitrary instruction location.
However, in the case where the granular behavior of a process has not been fully specified, we permit Common Lisp itself to shoulder its execution semantics, and it is difficult to jump to the middle of a Lisp-defined procedure.
To accommodate this, we essentially limit jumps to occur only at the end of each procedural step.
In effect, this becomes a form of continuation-passing style: each step in a procedure organizes only the code needed to continue its computation.}

Because \texttt{define-process-upkeep} describes a region of code which belongs to a particular active process, we again have the opportunity to define some utility macros.
\begin{description}
\item[\texttt{sync-receive}]
Busy-waits for a message of a specified type to arrive on a specified inbox.
Its syntax is identical to that of \receive.
\item[\texttt{with-replies}]
Gathers a family of responses, potentially from many different inboxes, by busy-waiting until they all arrive.
It then proceeds with the computation housed in its body.
\end{description}

\begin{example}
In \Cref{DOHRedux} we revise the example given in \Cref{DOHExample} to use the \texttt{process} framework, and in \Cref{FactorialServerEx} we extend it to service interprocess communication requests to perform a simple computation.
We also give a fresh example in \Cref{ColoringExample} of a distributed algorithm solving a standard problem in graph theory.

\begin{figure}[t]
\begin{minted}{cl}
(defclass processor (process)
 ()
 (:documentation "State of a simple CPU."))

(define-process-upkeep ((self processor) now) (HALT)
  (process-die))

(define-process-upkeep ((self processor) now) (PUSH value)
  (push value (process-data-stack self)))

(define-process-upkeep ((self processor) now) (MULI value)
  (push (* value (pop (process-data-stack self)))
        (process-data-stack self)))
        
(define-message-dispatch processor
  )
\end{minted}
\caption{Re-implementation of \Cref{DOHExample} using a \texttt{process}.
The \texttt{program-counter} and \texttt{instructions} slots have been absorbed into the command stack.}\label{DOHRedux}
\end{figure}

\begin{figure}[t]
\begin{minted}{cl}
(defclass arithmetic-server (processor) ())

;;; act like a server

(define-process-upkeep ((self arithmetic-server) now) (START)
  "This process sits in a loop, waiting to serve."
  (process-continuation self `(START)))

(define-process-upkeep ((self arithmetic-server) now) (EMIT address)
  "Generic instruction for emitting a message."
  (let ((result (pop (process-data-stack self))))
    (send-message address (make-message-rpc-done :result result))))

;;; factorial-specific functionality

(defstruct (message-factorial (:include message))
  "Message which requests a factorial computation."
  (n   nil :type (integer 0)))

(define-process-upkeep ((self arithmetic-server) now) (FACTORIAL n)
  "Computes n!, leaves result on the data stack."
  (cond
    ((zerop n)
     (process-continuation self `(PUSH 1)))
    (t
     (process-continuation self
                           `(FACTORIAL ,(1- n))
                           `(MULI ,n)))))

(define-message-handler handle-message-factorial
    ((self arithmetic-server) (message message-factorial) now)
  "Handles a inbound factorial request."
  (process-continuation self
                        `(FACTORIAL ,(message-factorial-n message))
                        `(EMIT ,(message-reply-channel message))))

;; advertises the factorial handler to externals
(define-message-dispatch arithmetic-server
  (message-factorial     'handle-message-factorial))
\end{minted}
\caption{Extends the \texttt{processor} defined in \Cref{DOHRedux} to allow it to serve requests for computing factorials.}\label{FactorialServerEx}
\end{figure}

\begin{figure}[t]
\begin{minted}{cl}
(defclass process-coloring (process)
  ((stopped?  ...)
  (color     ...)
  (neighbors ...)))

(defstruct (message-color-query (:include message)))

(define-rpc-handler handle-message-color-query
    ((process process-coloring)
     (message message-color-query)
     now)
  (process-coloring-color process))

(define-message-dispatch process-coloring
  (message-color-query 'handle-message-color-query))

(define-process-upkeep ((process process-coloring) now) (START)
  (process-continuation process `(QUERY)))
      
(define-process-upkeep ((process process-coloring) now) (QUERY)
  (when (process-coloring-stopped? process)
    (process-continuation process `(IDLE))
    (finish-with-scheduling))
  (let (listeners)
    (with-slots (color neighbors stopped?)
        process
      (process-continuation process `(QUERY))
      (setf color (random 3))
      (setf listeners
            (send-message-batch #'make-message-color-query
                                neighbors))
      (with-replies (replies) listeners
        (setf stopped? (not (member color replies)))))))))

(define-process-upkeep ((process process-coloring) now) (IDLE)
  (process-continuation process `(IDLE)))
\end{minted}
\caption{Randomized algorithm for 3-coloring a line which uses only local network structure~\cite{Linial,Suomela}. Leverages some features of the standard library discussed in \Cref{StdLibSection}.}\label{ColoringExample}
\end{figure}
\end{example}

\begin{remark}
There is no centralized directory service by which processes on a network can discover one another.
Instead, processes become aware of each others' existence in three ways:
\begin{enumerate}
    \item The user can directly intervene (e.g., before simulation start) and write a process's public address into the slot of another.
    \item When a process spawns a subprocess, it can record the subprocess's public address.
    \item Addresses can be sent over channels, so that processes can tell each other about addresses of which they're aware.
    (So, a programmer could write their own directory service.)
\end{enumerate}
\end{remark}

\section{A standard library}
\label{StdLibSection}

All of this scaffolding permits us to write actor-based distributed algorithms with relative ease.
Such algorithms follow certain standard patterns to enable easy interprocess coordination, and \aether provides built-in support for some of these patterns in the form of a standard library.

\subsection{Remote procedure calls}

One such pattern is the remote procedure call: one process queries one or more other processes, then awaits the result (or, at least, a signal that they have finished) before using it to continue its own computation.
This is typically implemented in terms of \aether primitives as:
\begin{enumerate}
    \item The caller registers an inbox, the ``private reply inbox''.%
    \footnote{%
    This automatically associates the reply message with the particular query which generated it, where we may otherwise have had to distinguish between several inbound replies at the process's public address.
    }
    \item The caller forms an RPC request with the private inbox listed as the return address and sends it to the public address of the remote process.
    \item The caller then performs a blocking receive, waiting to find a response in the private inbox.
    \item The remote process services the request, which triggers the desired computation.
    \item At the conclusion of the computation, the remote process prepares a \texttt{message-rpc-done}, which it sends back to the private address.
    \item Upon receiving that message, the caller unregisters the reply inbox and proceeds with its own computation.
\end{enumerate}

This involves some boilerplate code on both sides of the protocol, the caller and the callee.
On the side of the caller, we have:
\begin{description}
\item[\texttt{with-replies}]
Manages the blocking receive.
It also supports the automatic unboxing of the replies it receives (e.g., by mapping \texttt{\#'message-rpc-done-result} over the message objects), waiting on multiple mailboxes in parallel, and gracefully handling return-to-sender replies.
\end{description}
On the side of the callee, we have:
\begin{description}
\item[\texttt{define-rpc-handler}]
Extends \texttt{define-message-handler} by testing whether the return address of the inbound message is non-null and, if so, returning the result of its body computation via a \texttt{message-rpc-done} envelope.
\item[\texttt{define-message-subordinate}]
Defines a non-interrupting message handler.
A limitation of \texttt{define-message-handler} is that it functions akin to an interrupt: it cannot perform any operations that take nontrivial simulation time, and at its completion control passes back to the next frame on the call stack.
In order to perform computations which require simulation time (e.g., if they need to perform RPC calls of their own), the handler can install call stack frames of its own, interrupting whatever procedure was being worked on previously and blocking it from continuing until the call resolves.
As this blocking behavior can also be undesirable, \texttt{define-message-subordinate} provides an alternative: it spawns a parallel thread to handle the message, with all the features of \texttt{define-process-upkeep} available (e.g., \texttt{with-replies}).
\end{description}

\subsection{Broadcast and convergecast}

We also provide tools for common patterns in the domain of distributed graph algorithms.
In that context, processes correspond to problem graph nodes, and communication channels correspond to problem graph edges.
As an algorithm progresses, nodes communicate and form graph structures (e.g., rooted trees), and these structures may have to internally coordinate to take further action.
This coordination is conveniently accomplished via \textit{broadcast} and \textit{convergecast} protocols~\cite{GHS,Aspnes}.

The goal of a broadcast is to announce a message across a network, e.g., an instruction to enact some effect.
When a process receives a broadcast message, it performs an associated action, calculates the next set of targets (e.g., all adjacent nodes except the sender), and then forwards the message along to those targets.%
\footnote{%
We use the word ``broadcast'' to capture the recursive, potentially multi-target nature of this operation.
The message need not be forwarded to \emph{all} targets on the network.
}
Thus, a broadcast can be thought of as a way to recursively act on a distributed graph structure.
To facilitate this operation, we provide an extension to the standard message handler called \texttt{define-broadcast-handler}, which pushes a \texttt{BROADCAST} command onto the command stack before executing the body of the handler.
This command forwards the broadcasted message to the next round of targets, optionally awaiting their reply and acknowledging the broadcaster in the form of a \texttt{message-rpc-done}.

When a process wants to gather some information from across the network, it can do so using convergecast.
This protocol is similar to broadcast, with an additional step: each process reports the result of performing its action, and as reports are sent back to the convergecast originator, they are summarized by a user-specified computation in order to produce a final result.
As a trivial example, convergecast can be used to count the number of nodes in a distributed tree, where each node reports the number $1$ and the reports are summed as they are relayed back to the root.

\subsection{Locking}

Some process interactions are \textit{nonatomic}, in the sense that they might interleave with other process interactions.
Nonatomicity is why a distributed algorithm achieves a speedup over a sequential algorithm: a completely atomic system is essentially indistinguishable from a sequential system.
Nonetheless, atomicity also remains a valuable property, as it is often critical to the correctness of an algorithm.
One then wants to confer atomicity onto those sections of the algorithm which require it for correctness, called \textit{critical sections}, and to otherwise work nonatomically to take advantage of parallel process interactions.
Any mechanism by which we can confer controlled atomicity is called a \textit{lock}.
The \textit{broadcast lock}, a cousin of Lamport's mutual exclusion mechanism~\cite{Lamport}, and a special case of the more general broadcast operation described previously, is a particular form of locking which appeared often in our application work.
As such, we have included it as part of \aether's standard library.

To begin the protocol, the process which wishes to hold the lock calculates the set of clients it must lock and sends them each a lock request.
Each client either replies with a failure, or it gathers locks from its sub-clients, and so on.
When each client hears successful replies from each of its sub-clients, it opens a new, private inbox and sends that address to the lock holder as its signal of success.
Upon success, the lock holder then does its critical work; upon failure, it knows to avoid the critical work, perhaps to try again later.
In either case, it unwinds any client locks it did acquire by sending a finish message to the private addresses, then awaiting signals from each that they have finished.

The subclass \texttt{process-lockable} of \texttt{process} implements this algorithm within \aether.
The relevant hooks are as follows:
\begin{description}
\item[\texttt{BROADCAST-LOCK}] A command used to initiate lock requests.
\item[\texttt{message-lock-request}] A message type used to deliver lock requests, replete with message handlers.
\item[\texttt{process-lockable-targets}] A generic function used to calculate the clients on which to acquire recursive locks.
\end{description}

\begin{example}
In \Cref{LockingExample}, we provide an example of a pair of processes that use a broadcast lock to manage a race condition.
Several writer processes which want to transmit data to a single reading process first acquire a lock on it before transmitting their multi-part payloads, retrying if their lock is denied.
In this way, the reading process receives the payloads without interleaving, though they may arrive in any order relative to one another.

\begin{figure}
\begin{minted}{cl}
(defclass writer (process-lockable)
  ((transmit-list ...)
  (target        ...)))

(defclass reader (process-lockable)
  ((receive-list ...)))

(defstruct (message-write (:include message))
  payload)

(define-process-upkeep ((process reader) now) (START)
  (process-continuation process `(START)))

(define-process-upkeep ((process writer) now) (START)
  (unless (endp (writer-transmit-list process))
    (with-slots (target) process
      (process-continuation process
                            `(BROADCAST-LOCK (,target))
                            `(TRANSMIT)
                            `(BROADCAST-UNLOCK)
                            `(START))))))

(define-process-upkeep ((process writer) now) (TRANSMIT)
  (let ((next (pop (writer-transmit-list process))))
    (unless (or (process-lockable-aborting? process)
                (null next))
      (send-message (writer-target process)
                    (make-message-write :payload next))
      (process-continuation process `(TRANSMIT)))))

(define-rpc-handler handle-message-write
    ((process reader) (message message-write) now)
  (push (message-write-payload message)
        (reader-receive-list process)))

(define-message-dispatch writer
  )

(define-message-dispatch reader
  (message-lock  'handle-message-lock)
  (message-write 'handle-message-write))

(defmethod process-lockable-targets ((process reader))
  nil)
\end{minted}
\caption{Writer processes mutually exclude their transmission sequences to a commonly held reader process by first acquiring its lock.}\label{LockingExample}
\end{figure}
\end{example}

\section{Instrumentation}

Much of the value in writing simulation software is having a low-cost environment in which to do software development: one can implement complex software and verify at least some of its behaviors before testing it on a true physical system, which may itself be expensive to build.
As previously discussed, another large source of value comes from the opportunity to design the physical system itself: if the simulation is faithful enough to give an estimate of resource consumption, then its analysis can be used to set hardware performance requirements.
Toward both of these goals, \aether provides tools for analyzing the behavior of programs that use its simulation framework.

In sequel work, we describe an elaborate application of these tools to a hardware design problem in quantum computing.

\subsection{Logging}

A significant challenge in designing distributed software is diagnosing errors that arise: the effect of an error can be very far from its source, both in terms of position in the code base as well as in simulation state.
To help with this, \aether provides a structured logging service which aggregates log entries in the form of Lisp objects.
Many of the \aether primitives automatically emit these log messages, and \texttt{log-entry} allows the programmer to emit additional user-defined messages.
Because these messages are retained as Lisp objects, they are relatively easy to query and filter, allowing the programmer to ``zoom in'' on particular process interactions in a sea of other information.

\begin{figure}[t]
\includegraphics[width=\textwidth/3]{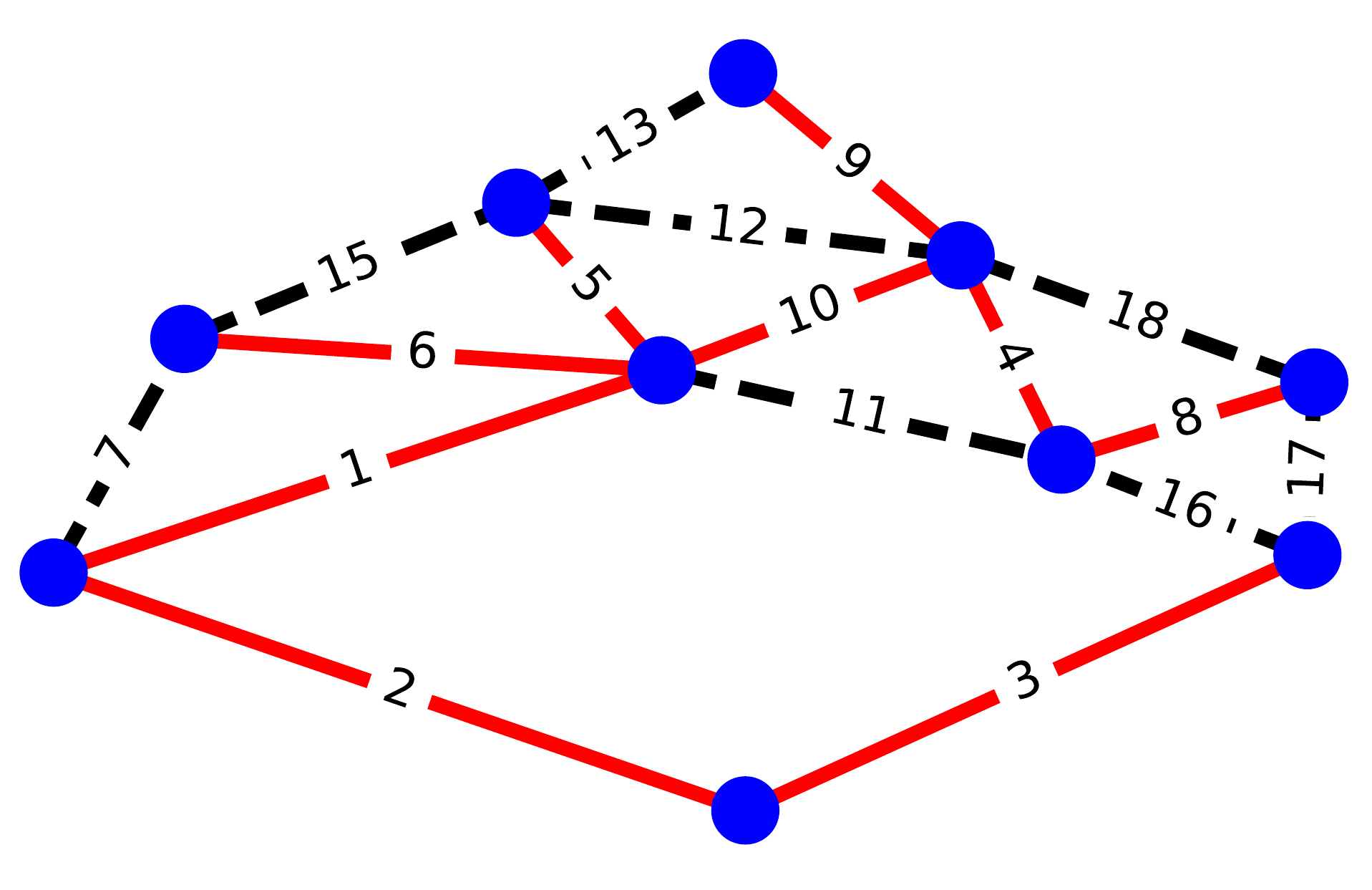}
\begin{minted}{cl}
MSG-REPORT:   21    MSG-ACCEPT:    7    MSG-REJECT:    9
MSG-TEST:     24    MSG-INITIATE: 12    MSG-CONNECT:  12
TOTAL:        85
\end{minted}
\caption{%
(top) Weighted problem graph (10 nodes, 17 edges) with minimum-weight spanning tree indicated by red solid edges.
(bottom) Output of the \texttt{print-message-report} logging mechanism when running the GHS algorithm~\cite{GHS} on the above graph.
In particular, we find a message total of 85, which is indeed less than the upper bound of 200.
}\label{MSTStats}
\end{figure}

\begin{example}
In \Cref{MSTStats} we provide an example of using some of the built-in logging facilities to verify that an implementation of the GHS algorithm runs using fewer messages than the predicted upper bound of $5 N \log N + 2 E$.
See \Cref{GHSAlgorithm} for  implementation details.
\end{example}

\subsection{Dereferencing}

Part of the conceit of distributed programming is that processes are unable to modify each others' states except by communication through agreed-upon protocols.
To enforce this within a simulated context, where a programmer could use the host language's ability to inspect objects directly, we make a concerted effort to retain only the public addresses of processes.
These are decoupled from the processes that own them, and hence they prevent their owner processes from being directly manipulated by the host language.

While necessary for the basic semantic validity of a distributed application, one frequently wants to dissolve exactly this barrier when debugging that application.
Diagnosing and tracing errors almost always requires inspection of the states of \emph{several} processes---but which processes can only be known at runtime, and even then they are known only by their public address.
We provide in \aether an optional debugging utility for marrying public addresses to their owners, enabling this critical form of state tracing.

\subsection{Call tracing}

In addition to understanding in detail the behavior of particular interactions, it is also useful to understand the aggregate statistics of interactions, as this has a direct impact on algorithm performance.
To this end, \aether provides a ``tracing'' mechanism that records a set of performance statistics for interprocess queries, e.g., number of targets in a batch send, or time spent waiting for a query to resolve (and, so, distinct from behavioral logging).
This, too, can be further instrumented to suit a user's needs.

\section{Avenues for improvement}

We close with some comments about the framework's shortcomings and indicate clear areas for improvement.

\subsection{Granular processor simulation}

While we have used the time-domain simulation features of an \aether-like library to emulate the behavior of a family of processors at the microarchitecture level, we have not done so in the same project where communication and high-level algorithms are relevant.
Because of this, we have not explored what a microarchitecture-specific DSL for algorithm specification looks like, having instead relied on Common Lisp to set process command semantics.
One would surely learn a lot by pursuing this, and it would probably motivate nontrivial changes in the definition of an \aether process, and perhaps to the framework generally.

Additionally, without a concrete specification of the ``compute'' power available to each process (even in an example!), it is impossible to faithfully simulate compute-bound behavior of systems.
Instead, the simulations with \aether described in this document tacitly assume that they are I/O-bound, so that the compute power does not matter relative to the network delay.

\subsection{Procedure delineation}

In current practice we break ``procedures'' into sequences of ``commands'' in a way that is pleasing to the conscious programmer, but unstructured and unenforced from within the framework.
It would be preferable to have a language-level method for delineating procedures, as well as for calling and returning from them.

Another side effect of manual manipulation of procedure bodies is that the validity of the command stack is not checked at compile time.
Whereas the Common Lisp compiler can check whether a static procedure call comports to its type signature, we have not provided \aether with the ability to warn a programmer whether the commands to be injected are defined at all, much less whether they match any advertised type signatures.

In a different but related direction, the locking system is not robust to malicious use.
Without further programming intervention, a locked process continues to respond to all messages, including in particular messages sent by actors other than the lock holder and messages which may belong to some other critical section.
Better segmentation of procedures is not a prerequisite for improving on these problems, but it would make that work simpler.

\subsection{A middleware DSL}

Apart from setting out the details of a microarchitecture, it is a separate problem to specify a useful DSL for writing distributed algorithms in an actor-based framework.
There are many obvious constraints to consider here, including flexibility, user-friendliness, and compiler-friendliness for mapping onto any potential microarchitecture embodiment.
We have only begun to describe such a DSL here in the form of the framework around \texttt{process}.

A less obvious and extremely promising aspect of designing a DSL is the possibility of introducing a type system suitable for analyzing the interactions of processes and enabling a compiler targeting a microarchitecture to reason about programs and optimize them.
A likely key starting point for such a project is a ``closed world'' hypothesis: process-based typing systems are difficult to govern when the processes can receive arbitrary queries from unknown third parties, but if the ingress points for such third parties are explicitly marked, initial progress seems possible.

\subsection{Granular network components}

Network specification is somewhat underexplored.
While it is currently very featureful---one can enact nontrivial routing strategies on nontrivial networks---it has not ever been pressured to become a ``faithful'' emulation of real networked electronics.
For instance, \aether presently measures the concept of network pressure by the packet count a node is processing, but the size and contents of those packets are not taken into account.
Serialization and deserialization strategies would increase the fidelity of this aspect of the simulation.
In a different direction, phenomena such as network noise (so: dropped packets, unpredictable delays, and so on) are not modeled.
Modifying the simulator to support this, modifying the messaging system to be robust against it, and exposing a networking layer where the user is not obligated to use the message-passing system would each be a valuable project.

\section*{Acknowledgements}

The authors owe debts of gratitude to several collaborators, friends, and family.
Separately and at different times, Charles Hadfield and Sergio Boixo both suggested that we take an interest in a particular form of topological quantum error correction, which ultimately led to this project and its sequels.
We are grateful to Colm Ryan for providing valuable guidance early in the course of this project, and to Erik Davis and Michael J.\ Fischer for feedback on a draft of this manuscript.
We would also like to thank Jade Lilitri for keeping our spirits ``up''.
Lastly, the first author would like to thank Samrita Dhindsa for all manner of support as this work was carried out.

\bibliography{main}

\appendix

\section{Implementation of GHS Algorithm}
\label{GHSAlgorithm}

Below we have an implementation of a canonical distributed algorithm using \aether's \texttt{process} framework. As indicated by the docstrings, each command or message handler definition maps directly to a section of the pseudocode algorithm in the appendix of the GHS paper~\cite{GHS}. The snippet is truncated for brevity---for the full implementation, please refer to the file \texttt{distributed-mst.lisp} in the \texttt{tests/examples} directory of the \aether repository~\cite{aether}.\\

\begin{minted}{cl}
(define-process-upkeep ((node fragment-node) now) (START)
  "Section 1: Response to spontaneous awakening."
  (process-continuation node `(START))
  (when (eql (slot-value node 'node-state) ':SLEEPING)
    (process-continuation node `(WAKEUP))))

(define-process-upkeep ((node fragment-node) now) (WAKEUP)
  "Section 2: Procedure WAKEUP."
  (let ((address (process-public-address node)))
    (with-slots (adjacent-edges find-count fragment-level) node
      (when (eql (slot-value node 'node-state) ':SLEEPING)
        (let ((minimum-weight-edge
                (find-minimum-weight-edge adjacent-edges)))
          (with-slots (edge-state)
              (gethash minimum-weight-edge adjacent-edges)
            (setf edge-state ':BRANCH find-count 0 fragment-level 0
                  (slot-value node 'node-state) ':FOUND)
            (send-message minimum-weight-edge
                          (make-msg-connect :edge address
                                            :level 0))))))))



(define-message-handler handle-msg-connect
    ((node fragment-node) (message msg-connect) time)
  "Section 3: Response to receipt of Connect(L) on edge j."
  (let ((address (process-public-address node)))
    (with-slots (fragment-level fragment-weight node-state) node
      (with-slots (edge level) message
        (with-slots (edge-state edge-weight)
            (gethash edge (slot-value node 'adjacent-edges))
          (cond
            ((< level fragment-level)
             (setf edge-state ':BRANCH)
             (send-message edge (make-msg-initiate
                                 :edge address
                                 :level fragment-level
                                 :state node-state
                                 :weight fragment-weight))
             (when (eql node-state ':FIND)
               (incf (slot-value node 'find-count))))
            ((eql edge-state ':BASIC)
             (send-message address message))
            (t
             (send-message edge (make-msg-initiate
                                 :edge address
                                 :level (1+ fragment-level)
                                 :state ':FIND
                                 :weight edge-weight)))))))))

(define-message-handler handle-msg-initiate
    ((node fragment-node) (message msg-initiate) time)
  "Section 4: Response to receipt of Initiate(L, F, S) on edge j."
  (let ((address (process-public-address node)))
    (with-slots (adjacent-edges find-count) node
      (with-slots (edge level state weight) message
        (setf (slot-value node 'best-edge)       nil
              (slot-value node 'best-weight)     most-positive-fixnum
              (slot-value node 'fragment-level)  level
              (slot-value node 'fragment-weight) weight
              (slot-value node 'in-branch)       edge
              (slot-value node 'node-state)      state)
        (let ((edges (find-specific-edges adjacent-edges
                                          :desired-state ':BRANCH
                                          :not-edge edge)))
          (flet ((payload-constructor ()
                   (make-msg-initiate :edge address :level level
                                      :state state :weight weight)))
            (send-message-batch #'payload-constructor edges))
          (loop :repeat (length edges) :when (eql state ':FIND)
                :do (incf find-count)))
        (when (eql state ':FIND)
          (process-continuation node `(TEST)))))))

(define-process-upkeep ((node fragment-node) now) (TEST)
  "Section 5: Procedure TEST."
  (with-slots (adjacent-edges test-edge) node
    (let ((minimum-weight-edge
            (find-minimum-weight-edge adjacent-edges
                                      :desired-state ':BASIC)))
      (setf test-edge minimum-weight-edge)
      (if minimum-weight-edge
          (send-message test-edge
                       (make-msg-test
                        :edge (process-public-address node)
                        :level (slot-value node 'fragment-level)
                        :weight (slot-value node 'fragment-weight)))
          (process-continuation node `(REPORT))))))

(define-message-handler handle-msg-test
    ((node fragment-node) (message msg-test) time)
  "Section 6: Response to receipt of Test(L, F) on edge j."
  (let ((address (process-public-address node)))
    (with-slots (adjacent-edges test-edge) node
      (with-slots (edge level weight) message
        (with-slots (edge-state) (gethash edge adjacent-edges)
          (cond
            ((> level (slot-value node 'fragment-level))
             (send-message address message))
            ((/= weight (slot-value node 'fragment-weight))
             (send-message edge (make-msg-accept :edge address)))
            (t
             (when (eql edge-state ':BASIC)
               (setf edge-state ':REJECTED))
             (if (or (null test-edge) (not (address= test-edge edge)))
                 (send-message edge (make-msg-reject :edge address))
                 (process-continuation node `(TEST))))))))))

(define-message-handler handle-msg-accept
    ((node fragment-node) (message msg-accept) time)
  "Section 7: Response to receipt of Accept on edge j."
  (with-slots (adjacent-edges best-edge best-weight test-edge) node
    (with-slots (edge) message
      (with-slots (edge-weight) (gethash edge adjacent-edges)
        (setf test-edge nil)
        (when (< edge-weight best-weight)
          (setf best-edge edge best-weight edge-weight))
        (process-continuation node `(REPORT))))))

(define-message-handler handle-msg-reject
    ((node fragment-node) (message msg-reject) time)
  "Section 8: Response to receipt of Reject on edge j."
  (with-slots (adjacent-edges) node
    (with-slots (edge) message
      (with-slots (edge-state) (gethash edge adjacent-edges)
        (when (eql edge-state ':BASIC) (setf edge-state ':REJECTED))
        (process-continuation node `(TEST))))))

(define-process-upkeep ((node fragment-node) now) (REPORT)
  "Section 9: Procedure REPORT."
  (with-slots (best-weight find-count in-branch test-edge) node
    (when (and (zerop find-count) (null test-edge))
      (setf (slot-value node 'node-state) ':FOUND)
      (send-message in-branch (make-msg-report
                               :edge (process-public-address node)
                               :weight best-weight)))))

(define-message-handler handle-msg-report
    ((node fragment-node) (message msg-report) time)
  "Section 10: Response to receipt of Report(w) on edge j."
  (with-slots (best-edge best-weight find-count in-branch) node
    (with-slots (edge weight) message
      (cond
        ((not (address= edge in-branch))
         (decf find-count)
         (when (< weight best-weight)
           (setf best-weight weight best-edge edge))
         (process-continuation node `(REPORT)))
        ((eql (slot-value node 'node-state) ':FIND)
         (send-message (process-public-address node) message))
        ((> weight best-weight)
         (process-continuation node `(CHANGE-ROOT)))
        ((and (= weight best-weight) (= weight most-positive-fixnum))
         (process-continuation node `(HALT)))))))

(define-process-upkeep ((node fragment-node) now) (CHANGE-ROOT)
  "Section 11: Procedure CHANGE-ROOT."
  (let ((address (process-public-address node)))
    (with-slots (adjacent-edges best-edge fragment-level) node
      (with-slots (edge-state) (gethash best-edge adjacent-edges)
        (cond
          ((eql edge-state ':BRANCH)
           (send-message best-edge
                         (make-msg-change-root :edge address)))
          (t
           (send-message best-edge
                         (make-msg-connect :edge address
                                           :level fragment-level))
           (setf edge-state ':BRANCH)))))))

(define-message-handler handle-msg-change-root
    ((node fragment-node) (message msg-change-root) time)
  "Section 12: Response to receipt of Change-root."
  (process-continuation node `(CHANGE-ROOT)))
\end{minted}

\end{document}